\documentclass[12pt]{article}
\usepackage{amsmath} 
\usepackage{pstricks}
\usepackage{longtable}
\usepackage[dvips]{graphicx}       
\usepackage{color}
\usepackage{cite}
\usepackage{amssymb}
\input{epsf} 
\newcommand\as{\alpha_{\mathrm{S}}} 
\newcommand\f[2]{\frac{#1}{#2}}

\def\beq{\begin{equation}} 
\def\eeq{\end{equation}} 
\def\to{\rightarrow} 
\def\nn{\nonumber}

\def\b0{\beta_0}

\def\beeq{\begin{eqnarray}}
\def\eeeq{\end{eqnarray}}

\def\mur{\mu_R} 
\def\muf{\mu_F}
\def\mur2{\mu_R^2} 
\def\muf2{\mu_F^2}
\newcommand\dd{{\cal D}}

\newcommand {\apgt} {\ {\raise-.5ex\hbox{$\buildrel>\over\sim$}}\ }
\newcommand {\aplt} {\ {\raise-.5ex\hbox{$\buildrel<\over\sim$}}\ }

\begin{document}

\begin{titlepage}
\renewcommand{\thefootnote}{\fnsymbol{footnote}}
\begin{flushright}
\end{flushright}
\par \vspace{-5mm}

\begin{center}
{\Large \bf

Next-to-Next-to-Leading Order QCD Corrections in Models of TeV-Scale Gravity

}
\end{center}
\par \vspace{-1mm}
\begin{center}
{\bf Daniel de Florian}\footnote{deflo@df.uba.ar}$^{(a)}$,
{\bf Maguni Mahakhud}\footnote{maguni@imsc.res.in}$^{(b)}$,
{\bf Prakash Mathews}\footnote{prakash.mathews@saha.ac.in}$^{(c)}$,\\
[5pt]
{\bf Javier Mazzitelli}\footnote{jmazzi@df.uba.ar}$^{(a)}$ and
{\bf V. Ravindran}\footnote{ravindra@imsc.res.in}$^{(b)}$\\

\vspace{10mm}

$^{(a)}$ Departamento de F\'\i sica, FCEyN, Universidad de Buenos Aires, \\
(1428) Pabell\'on 1, Ciudad Universitaria, Capital Federal, Argentina

$^{(b)}$ The Institute of Mathematical Sciencies\\
C.I.T Campus, 4th Cross St, Tharamani Chennai,\\
Tamil Nadu 600 113, India

$^{(c)}$ Saha Institute of Nuclear Physics, 1/AF Bidhan Nagar, Kolkata 700 064, India

\vspace{5mm}

\end{center}

\par \vspace{2mm}
\begin{center} {\large \bf Abstract} \end{center}
\begin{quote}
\pretolerance 10000

We compute the next-to-next-to-leading order QCD corrections to the graviton production in models of TeV-scale gravity, within the soft-virtual approximation.

For the Arkani-Hamed, Dimopoulos and Dvali (ADD) model we evaluate the contribution to the Drell-Yan cross section, and 
we present distributions for the di-lepton invariant mass at the LHC with a center-of-mass energy $\sqrt{s_H}=14\text{ TeV}$.
We find a large $K$ factor ($K\simeq1.8$) for large values of invariant mass, which is the region where the ADD graviton contribution dominates the cross section.
The increase in the cross section with respect to the previous order result is larger than $10\%$ in the same invariant mass region.
We also observe a substantial reduction in the scale uncertainty.

For the Randall-Sundrum (RS) model we computed the total single graviton production cross section at the LHC. We find an increase between $10\%$ and $13\%$ with respect to the next-to-leading order prediction, depending on the model parameters.
We provide an analytic expression for the NNLO $K$ factor as a function of the lightest RS graviton mass.

\end{quote}

\vspace*{\fill}
\begin{flushleft}
\end{flushleft}
\end{titlepage}

\setcounter{footnote}{1}
\renewcommand{\thefootnote}{\fnsymbol{footnote}}

\section{Introduction}

Several ideas have been put forth over the years to address the gauge
hierarchy problem, one alternate approach is to postulate the existence
of extra spatial dimensions wherein only gravity is allowed to propagate.
Depending on the geometry of the extra dimension there are different
scenarios {\em viz.}\ (i) large extra dimensions (ADD) \cite{ADD} and
(ii) warped extra dimensions models (RS) \cite{RS1}.

In the ADD model there are $d$ flat large extra spatial dimensions of
same radii $R$, compactified on a $d$-dimensional torus.  Due to the
larger volume of extra dimension available for gravity, it appears weak
in the 4-dimensions where the SM particles and their interactions are
restricted to.  By Gauss's law the $4+d$ dimensional fundamental Planck
scale gets related to $4$-dimensional Planck scale $(M_P)$ {\em via}
the volume factor $(R^d)$ of extra dimension and a large enough volume
could result in the fundamental Planck scale of the order of a TeV,
there by ensuring the resolution of the hierarchy problem.  Propagation
of gravity in the large extra dimensions results in a continuous spectrum
of Kaluza-Klein (KK) modes in 4-dimensions with small mass splitting 
of the order of $1/R$.

The interaction of the spin-2, KK modes $(h_{\mu\nu})$ with the SM particles
is {\em via} the energy momentum tensor $T^{\mu\nu}$ of the SM, which is
universally suppressed by a coupling $\kappa=\sqrt{16 \pi}/M_P$,  
\begin{eqnarray}
{\cal L}_{ADD} = - \frac{\kappa}{2} \sum_{\vec n} T^{\mu \nu} (x) 
~h_{\mu \nu}^{(\vec n)} (x) .
\end{eqnarray}
Two types of process involving graviton are possible {\em viz.}\ exchange of
virtual graviton and production of a real graviton.  For processes involving
virtual KK mode exchange between SM particles, summation over the high multiplicity
KK modes leads to a compensation of the $\kappa$ suppression.  Due to the
continuous KK modes spectrum, the summation is replaced by an integral
with appropriate density of state $\rho (m_{\vec n})$ of KK modes \cite{HLZ}.
ADD model being an effective low energy theory, the integral is cutoff at a
scale $M_S$ that defines the onset of quantum gravity.  The cross section could hence
be appreciable at collider energies, giving rise to non-resonant enhancement of
the high invariant mass regions of a di-final state production
\cite{HLZ,GRW,di-final} or final
states involving more particles \cite{tri-final}.  

Real graviton production leads to missing energy signal and a cross section for
the production of a single graviton $d \sigma_{m_{\vec n}}$ has to be convoluted
with the graviton density of state to get the inclusive cross section.
Here too the collective contribution of the KK modes results in observable effects
at the collider.

Next to leading order QCD corrections are available for most of the di-final
state process in the ADD model {\em viz.}\ $\ell^+ \ell^-$ \cite{di-ll1,di-ll2,di-ll3}, 
$\gamma \gamma$ \cite{di-ph1,di-ph2} and $ZZ$ \cite{di-ZZ1,di-ZZ2} and $W^+ W^-$
\cite{di-WW1,di-WW2},
in addition these processes have been extended to NLO+PS accuracy
\cite{di-ph+ps,di-final+ps}.  In the case of missing energy signals, as a 
result of a real graviton in association with (a) jet \cite{jEt}, (b) photon
\cite{phEt} and (c) electro-weak gauge boson \cite{ZWEt} have been studied to
NLO in QCD.  The $K$ factors for these process at the LHC are large.  It is
important to study the full impact of QCD corrections in terms of the shape
of the various distributions and not just an overall normalising $K$ factor.

The RS model is an alternate extra dimension model with one exponentially warped
extra dimension $y$ with radius of compactification $r_c$, where again only gravity
is allowed to propagate.  In this model there are two 3-branes; gravity resides on
the Planck brane at $y=0$ and it appear weaker on the TeV brane located at $y=\pi r_c$
due to the exponential warping.  A mass scale on the TeV brane $\Lambda_{\pi} = 
{\overline M}_P \exp (-k \pi r_c)$ as a result of gravity resides on the Planck
brane could be of the order of a TeV, for $k r_c \sim 12$.  $k$ is the 
curvature of the extra dimension.  The interaction Lagrangian of the RS KK mode
with the SM particles are given by 
\begin{eqnarray}
{\cal L}_{RS} = - \frac{1}{\Lambda_{\pi}} \sum_{n=1}^\infty T^{\mu \nu} (x) 
~ h_{\mu \nu}^{(n)} (x) .
\end{eqnarray}
The zero mode corresponding to the massless graviton which is $M_P$ suppressed 
is not included in the sum.  As a result of the warped geometry of the extra
dimension the characteristic mass spectrum of the KK modes is $M_n = x_n k \exp 
(-k \pi r_c)$, where $x_n$ are the zeros of the Bessel's function.  In the RS
case, the resonant production of KK modes would be observed in a pair production
of final state SM particles.

In the RS model, NLO QCD calculation to various processes for the resonant production
of a RS graviton has been done at the LHC for di-lepton production \cite{di-ll2,di-ll3},
di-photon production \cite{di-ph2}, di-neutral electroweak gauge boson production
\cite{di-ZZ2} and charged electroweak gauge boson production $W^+ W^-$ \cite{di-WW2}.
Various distributions have also been considered and the $K$ factors are large in 
the resonant region where the gravity effects are large.

With the inclusion of the full NLO computation the theoretical uncertainties are
reduced when going form LO to NLO, but for most of these processes the renormalisation scale dependence begins at the NLO level, and the total theoretical uncertainties are still large at this order.
Furthermore, the size of the NLO QCD corrections makes it necessary to reach higher orders in the perturbative series to be able to provide accurate predictions.

A full NNLO calculation requires the evaluation of the double real radiation, real emission from one-loop corrections and the pure virtual two-loop amplitudes.
However, the dominant terms are given by the soft and virtual contributions, which can be obtained in a simpler way.
This fact is a general feature of the production of a large invariant mass system in hadronic collisions. Since parton distributions grow fast for small fractions of the hadron momentum, the partonic center-of-mass energy tends to be close to the system invariant mass, and the remaining energy only allows for the emission of soft particles.
For this reason, the soft-virtual (SV) approximation is expected to be accurate for a large number of processes.

In this work we compute the NNLO QCD corrections to the graviton production at the LHC for the ADD and RS models, within the soft-virtual approximation.
The paper is organized as follows.
In section 2 we present the partonic cross sections for di-lepton production in the ADD model and for single graviton production in the RS model.
In section 3 we analyse the phenomenological results for the LHC.
Finally, in section 4 we present our conclusions.




\section{Partonic Cross Section}

As mentioned before, the spin-2 form factor has been calculated recently by us in Ref. \cite{deFlorian:2013sza}.
Using those results, the complete two-loop corrections for single graviton production and di-lepton production mediated by a graviton can be obtained, for both gluon-gluon and quark-antiquark partonic subprocesses.
These contributions include the interference between the two-loop and the tree-level amplitudes and the square of the one-loop amplitudes.
These results, computed within the dimensional regularization scheme, are of course divergent in the limit $n\to 4$, being $n$ the space-time dimension. To obtain a finite and physically meaningful quantity we have to add the corresponding real corrections, which cancel the infrared divergences.

On the other hand, in Ref. \cite{deFlorian:2012za} some of us derived a universal formula for the NNLO inclusive cross section of any colourless final state process within the soft-virtual approximation.
This formula depends on the particular process only through an infrared regulated part of the one and two-loop corrections, which can be obtained from the full virtual result (see Ref. \cite{deFlorian:2012za} for more details).
In this way we can obtain the NNLO corrections to single graviton production and gravity mediated di-lepton production within the soft-virtual approximation.
We have also obtained the NNLO-SV result by summing explicitly the soft contributions of Refs. \cite{Ravindran:2005vv,Ravindran:2006cg,Catani:2003zt,Moch:2005ky,Laenen:2005uz,Idilbi:2005ni}, arriving to the same results.

We provide here the final results, including the previous orders contributions. For the sake of brevity, we refer the reader to Ref. \cite{Hamberg:1990np,Harlander:2002wh} for the SM contribution to the di-lepton production cross section.
We remark that, as it was already noticed in Ref. \cite{di-ll1}, the interference between SM and gravity contribution to the di-lepton production invariant mass distribution identically vanishes.

We begin with the ADD model.
The graviton contribution to the di-lepton invariant mass ($Q$) distribution at the parton level can be cast in the following way:
\beeq
\f{d\hat\sigma}{dQ^2}=
{\cal F}_{\text{ADD}} \,
z \,
\Delta_{ab}(z)\,,
\eeeq
where $z=Q^2/s$, being $s$ the partonic center-of-mass energy, and $a, b$ denote the type of massless partons ($a, b=g,q,\bar q$, with $n_f$ different flavours of light quarks).
The constant ${\cal F}_{\text{ADD}}$ takes the following form:
\beq
{\cal F}_{\text{ADD}}=
\f{\kappa^4 Q^4}
{640\pi^2}
\left\vert
{\cal D}(Q^2)\right\vert^2\,,
\eeq
where the function ${\cal D}(Q^2)$ can be expressed as \cite{HLZ}
\beq
{\cal D}(Q^2)=16\pi\left(\f{Q^{d-2}}{\kappa^2 M_S^{d+2}}\right)I\left(\f{M_S}{Q}\right)\,.
\eeq
The integral $I$ is regulated by an ultraviolet cutoff, presumably of the order of $M_S$ \cite{GRW,HLZ}. This sets the limit on the applicability of the effective theory (for the di-lepton production this consistency would imply $Q<M_S$).
The summation over the non-resonant KK modes yields
\begin{eqnarray}
I(\omega)&=&- \sum_{k=1}^{d/2-1} {1 \over 2 k} \omega^{2 k}
 -{1 \over 2} \log(\omega^2-1)\,, \qquad \qquad \qquad  d={\rm even}\,,
\label{eq16}
\end{eqnarray}
\begin{eqnarray}
I(\omega)&=&
-\sum_{k=1}^{(d-1)/2} {1\over 2 k-1} \omega^{2 k-1}
 +{1 \over 2} \log \left({\omega+1}\over{\omega-1}\right) \,,
\quad \quad  d={\rm odd}\,.
\label{eq17}
\end{eqnarray}
On the other hand, for the RS model we have for the single graviton production cross section the following expression:
\beq
\hat\sigma=
{\cal F}_{\text{RS}} \,
z \,
\Delta_{ab}(z)\,,
\eeq
where the constant ${\cal F}_{\text{RS}}$ takes the following form:
\beq
{\cal F}_{\text{RS}}=\f{1}{\Lambda_\pi^2}\,.
\eeq
Notice that in this case we have $z=M_1^2/s$.

The coefficient function $\Delta_{ab}(z)$, which is independent of the model considered, has a perturbative expansion in terms of powers of the QCD renormalized coupling $\as$:
\beq
\Delta_{ab}(z) =
\sum_{i=0}^{\infty}
\left(\f{\as}{2\pi}\right)^i \Delta_{ab}^{(i)}(z)\,.
\eeq
At LO we only have nonzero contributions from $ab=gg$ and $ab=q\bar q$ (always equal to $ab=\bar q q$), which take the following form: 
\beeq
\Delta^{(0)}_{q \bar q}&=&\f{\pi}{8N_c}\delta(1-z)\,,\\[1ex]
\Delta^{(0)}_{gg}&=&\f{\pi}{2(N_c^2-1)}\delta(1-z)\,.
\eeeq
Here $N_c$ stands for the number of quark colors ($N_c=3$).
The NLO contributions, which have been calculated in Ref. \cite{di-ll1}, can be written in the following way:
\beeq
\Delta^{(1)}_{q \bar q}&=&\left({\pi \over 8 N_c}\right) C_F
\Bigg[ \Big(-10+4 \zeta_2\Big)\delta(1-z)+
4\, {\cal D}_0 \ln\left({Q^2 \over \mu_F^2}\right)
+8\, {\cal D}_1
 \\[1ex]
&&
+3 \delta(1-z)
\ln\left({Q^2 \over \mu_F^2}\right)
-2(1+z) \ln\left({Q^2 (1-z)^2 \over \mu_F^2 z}\right)
-4 {\ln(z) \over 1-z}+ {8 \over 3 z} - {8 z^2 \over 3}\Bigg]\,,
\nonumber \\[1ex]
\Delta^{(1)}_{q (\bar q) g}&=&\left({\pi \over 8 N_c}\right) 
\Bigg[ (-\f{7}{2}+{4 \over z} + z + z^2) \ln\left({Q^2 (1-z)^2 \over 
\mu_F^2 z}\right)
+\f{9}{4}-{3 \over z} +\f{9}{2} z-\f{7}{4} z^2 \Bigg]\,,
\eeeq
\beeq
\Delta^{(1)}_{g g}&=&\left({\pi \over 2 (N_c^2-1)}\right) C_A
\Bigg[ \Big(-{203 \over 18}+4 \zeta_2\Big)\delta(1-z)+
4\, {\cal D}_0 \ln\left({Q^2 \over \mu_F^2}\right)
+8\, {\cal D}_1
 \\[1ex]
&&
+{11 \over 3} \delta(1-z)
\ln\left({Q^2 \over \mu_F^2}\right)
+4 (-2+{1\over z}+z-z^2) 
\ln\left({Q^2 (1-z)^2 \over \mu_F^2 z}\right)
-4 {\ln(z) \over (1-z)}
\nonumber \\[1ex]
&&
-1-{11 \over 3 z}+ z+{11 z^2 \over 3}\Bigg]
+\left({\pi \over 2 (N_c^2-1)}\right) n_f \Bigg[\left({35 \over 18}-
{2 \over 3} \ln\left({Q^2 \over \mu_F^2}\right) \right) \delta(1-z)\Bigg]\,.\nn
\eeeq
Here $\mu_F$ and $\mu_R$ stand for the factorization and renormalization scales, and the $SU(N_c)$ Casimir operators are $C_F=\f{N_c^2-1}{2N_c}$ and $C_A=N_c$.
We have also defined the distributions ${\cal D}_i$ as
\beq
{\cal D}_i=\left(\f{\ln^i(1-z)}{1-z}\right)_+\,,
\eeq
where the $+$ symbol indicates the usual plus-prescription:
\beq
\int_0^1 dz\, f_+(z)\, g(z) = \int_0^1 dz\, f(z)\, [g(z)-g(1)]\,.
\eeq
The Riemann zeta function is denoted by $\zeta_i\equiv\zeta(i)$.

We present below the NNLO results in the soft-virtual approximation. Within this approximation we have only contributions to the gluon-gluon and quark-antiquark subprocesses, since the terms proportional to $\delta(1-z)$ and ${\cal D}_i$ (which are the ones we obtain within the SV approximation) are absent in other channels.
The result for the quark-antiquark subprocess is the following:
\beeq
\Delta^{(2)SV}_{q \bar q} \!\!\!\!&=&
\left(\f{\pi}{8N_c}\right)C_F^2\bigg\{
\bigg[
\f{2293}{48}-\f{35}{2}\zeta_2-31\zeta_3+\zeta_4
+\left(
-\f{117}{4}+6\zeta_2+44\zeta_3
\right)\ln\left(\f{Q^2}{\mu_F^2}\right)
\\[1ex]
&+&
\left(
\f{9}{2}-8\zeta_2
\right)\ln^2\left(\f{Q^2}{\mu_F^2}\right)
\bigg]\delta(1-z)
+64\zeta_3 \dd_0 -(80+32\zeta_2)\dd_1 +32\dd_3
\nn\\[1ex]
&-&
 8 \left(\dd_0 \left(2 \zeta _2+5\right)-3 \left(\dd_1+2 \dd_2\right)\right) \ln \left(\frac{Q^2}{\mu _F^2}\right)
+4\left(3 \dd_0+4 \dd_1\right) \ln^2\left(\f{Q^2}{\mu_F^2}\right)
\bigg\}
\nn\\[1ex]&+&
\left(\f{\pi}{8N_c}\right)C_A C_F\bigg\{
\bigg[
-\f{5941}{144}+\f{82}{9}\zeta_2+23\zeta_3-\f{3}{2}\zeta_4
+ \left(\f{22}{3} \zeta _2-6 \zeta _3+\f{17}{12}\right) \ln \left(\frac{\mu _R^2}{\mu _F^2}\right)
\nn\\[1ex]&+&
\f{11}{4} \ln ^2\left(\frac{\mu _R^2}{\mu _F^2}\right)+\f{1}{4} \ln \left(\frac{\mu _R^2}{Q^2}\right) \left(24 \zeta _3-11 \ln
   \left(\frac{\mu _R^2}{Q^2}\right)-79\right)
\bigg]\delta(1-z)
\nn\\[1ex]&+&
\left(
-\f{404}{27}+\f{44}{3}\zeta_2+14\zeta_3
\right)\dd_0
+\left(
\f{268}{9}-8\zeta_2
\right)\dd_1
-\f{44}{3}\dd_2
-\frac{44}{3} \dd_1 \ln
   \left(\frac{Q^2}{\mu_R^2}\right)
   \nn\\[1ex]
   &+&
\dd_0 \ln \left(\frac{Q^2}{\mu_F^2}\right) \left[\frac{11}{3} \ln \left(\frac{\mu_R^4}{\mu_F^2
   Q^2}\right)-4 \zeta_2+\frac{134}{9}\right]
\bigg\}\nn
\eeeq
\beeq
\hspace{1.5cm}&+& 
\left(\f{\pi}{8N_c}\right)C_F n_f\bigg\{
\bigg[
\f{461}{72}-\f{16}{9}\zeta_2+2\zeta_3
+
 \f{1}{2} \ln \left(\frac{\mu _R^2}{Q^2}\right) \left(\ln \left(\frac{\mu _R^2}{Q^2}\right)+7\right)
\nn\\[1ex]&-&
\f{1}{6}\ln \left(\frac{\mu _R^2}{\mu _F^2}\right) \left(8 \zeta _2+3 \ln \left(\frac{\mu _R^2}{\mu
   _F^2}\right)+1\right)
\bigg]\delta(1-z)
+
\left(
\f{56}{27}-\f{8}{3}\zeta_2
\right)\dd_0
-\f{40}{9}\dd_1+\f{8}{3}\dd_2
\nn\\[1ex]&+&
\frac{1}{9} \left[-2 \dd_0 \ln \left(\frac{\mu _R^2}{\mu _F^2}\right) \left(3 \ln \left(\frac{\mu _R^2}{\mu _F^2}\right)+10\right)+6 \dd_0 \ln ^2\left(\frac{\mu _R^2}{Q^2}\right)+4 \left(5 \dd_0-6
   \dd_1\right) \ln \left(\frac{\mu _R^2}{Q^2}\right)\right]
\bigg\}\nn\,.
\eeeq
On the other hand, the gluon-gluon contribution to the NNLO-SV partonic cross section takes the following form:
\beeq
\Delta^{(2)SV}_{gg} \!\!\!\!&=&
\left({\pi \over 2 (N_c^2-1)}\right)C_A^2 
\bigg\{
\bigg[
\f{7801}{1296}-\f{56}{9}\zeta_2-\f{22}{3}\zeta_3-\f{1}{2}\zeta_4
-\ln \left(\frac{Q^2}{\mu_R^2}\right) \left(\frac{121}{18}
   \ln \left(\frac{Q^2}{\mu_F^2}\right)+\frac{22 \zeta_2}{3}-\frac{2233}{108}\right)
\nn\\[1ex]
&+&
\ln \left(\frac{Q^2}{\mu_F^2}\right) \left(\left(\frac{121}{12}-8 \zeta_2\right) \ln
   \left(\frac{Q^2}{\mu_F^2}\right)+\frac{44 \zeta_2}{3}+38
   \zeta_3-\frac{1945}{54}\right)\bigg]\delta(1-z)
\\[1ex]
&+&
\left(
-\f{404}{27}+\f{44}{3}\zeta_2+78\zeta_3
\right)\dd_0
+\left(
-\f{544}{9}-40\zeta_2
\right)\dd_1
-\f{44}{3}\dd_2 + 32\dd_3
\nn\\[1ex]&+&
\ln \left(\frac{Q^2}{\mu_F^2}\right) \left[\left(\frac{55 \dd_0}{3}+16 \dd_1\right)
   \ln \left(\frac{Q^2}{\mu_F^2}\right)-\dd_0 \left(20
   \zeta_2+\frac{272}{9}\right)+\frac{88 \dd_1}{3}+48 \dd_2\right]\nn\\[1ex]
   &-&\ln
   \left(\frac{Q^2}{\mu_R^2}\right) \left(\frac{22}{3} \dd_0 \ln
   \left(\frac{Q^2}{\mu_F^2}\right)+\frac{44 \dd_1}{3}\right)
\bigg\}
\nn\\[1ex]
&+&
\left({\pi \over 2 (N_c^2-1)}\right)C_A n_f
\bigg\{
\bigg[
-\f{2983}{648}-\f{47}{18}\zeta_2+\f{16}{3}\zeta_3
+\ln \left(\frac{Q^2}{\mu_R^2}\right) \left(\frac{22}{9} \ln \left(\frac{Q^2}{\mu_F^2}\right)+\frac{4
   \zeta_2}{3}-\frac{791}{108}\right)
   \nn\\[1ex]
&-& 
\ln \left(\frac{Q^2}{\mu_F^2}\right) \left(\frac{11}{3}
   \ln \left(\frac{Q^2}{\mu_F^2}\right)+\frac{8 \zeta_2}{3}-\frac{719}{54}\right)
\bigg]\delta(1-z)
\nn\\[1ex]
&+&
\left(
\f{56}{27}-\f{8}{3}\zeta_2
\right)\dd_0
+\f{100}{9}\dd_1 + \f{8}{3}\dd_2
+
\frac{4}{3} \ln \left(\frac{Q^2}{\mu_R^2}\right) \left[\dd_0 \ln
   \left(\frac{Q^2}{\mu_F^2}\right)+2 \dd_1\right]
\nn\\[1ex]
&-&
\frac{2}{9} \ln \left(\frac{Q^2}{\mu_F^2}\right)
   \left[15 \dd_0 \ln \left(\frac{Q^2}{\mu_F^2}\right)-25 \dd_0+24 \dd_1\right]
\bigg\}
\nn\\[1ex] 
&+&
\left({\pi \over 2 (N_c^2-1)}\right)n_f^2
\bigg\{
\f{1225}{1296}+\f{2}{3}\zeta_2
+
 \f{1}{27} \ln \left(\frac{\mu _F^2}{Q^2}\right) \left[9 \ln \left(\frac{\mu _F^2}{Q^2}\right)-6 \ln \left(\frac{\mu _R^2}{Q^2}\right)+35\right]
\nn\\[1ex]
&-&\f{35}{54} \ln \left(\frac{\mu _R^2}{Q^2}\right)
\bigg\}\delta(1-z)
+
\left({\pi \over 2 (N_c^2-1)}\right)C_F n_f
\left[
\f{61}{12}-4\zeta_3+\ln\left(\f{\mu _F^2}{Q^2}\right)
\right]\delta(1-z)\nn\,.
\eeeq

These expressions are obtained by keeping only the most divergent terms of the real contributions when $z\to 1$, or equivalently, by keeping only the $\delta(1-z)$ and ${\cal D}_i$ distributions in the final result. However, the soft limit can be defined in a more natural way by working in Mellin (or $N$-moment) space, where instead of distributions in $z$ the dominant contributions are given by continuous functions of the variable $N$.
In fact, it was shown that large subleading terms arise when one attempts to formulate the soft-gluon resummation in $z$-space, and then all-order resummation cannot be systematically defined in $z$-space \cite{Catani:1996yz}. 
Also, in Refs. \cite{deFlorian:2012za,Catani:2003zt} it was shown that the soft-virtual approximation yields better results at NLO and NNLO for Higgs boson production and the Drell-Yan process if defined in $N$-space.

We will therefore work within the $N$-space formulation, in which we take the Mellin transform of the coefficient function $\Delta_{ab}(z)$ and drop all those terms that vanish when $N\to\infty$, which is the Mellin space analogous of $z\to 1$. For more details, see for example Ref. \cite{deFlorian:2012za}.



\section{Phenomenological Results}

\subsection{ADD Model}

In this section we provide the phenomenological results for the LHC, for a center-of-mass energy $\sqrt{s_H}=14\text{ TeV}$.
Taking into account the bounds on $M_S$ for different extra dimensions $d$ obtained by ATLAS \cite{ATLAS:2011ab} and CMS \cite{Chatrchyan:2011fq} collaborations, we choose for our present analysis the following values: $M_S=3.7\text{ TeV }(d=2)$, $3.8\text{ TeV }(d=3)$, $3.2\text{ TeV }(d=4)$, $2.9\text{ TeV }(d=5)$ and $2.7\text{ TeV }(d=6)$. 
We remark that for the SM contribution to the di-lepton production cross section at NNLO we always use the exact result.
On the other hand, for the soft-virtual approximation, used only in the NNLO graviton contributions, we always use the Mellin space definition.

To obtain the hadronic cross section we need to convolute the partonic result with the parton distribution functions (PDFs) in the following way:
\beq
\f{d\sigma}{dQ^2}(s_H,Q^2)=
\sum_{a,b}\int_0^1 dx_1 dx_2 f_{a/h_1}(x_1,\mu_F^2) f_{b/h_2}(x_2,\mu_F^2)
\int_0^1 dz\; \delta\!\left(z-\f{\tau}{x_1 x_2}\right) \f{d\hat\sigma_{ab}}{dQ^2}(s,Q^2)\,,
\eeq
where $s_H$ is the hadronic center-of-mass energy, and $\tau=Q^2/s_H$.
In all cases we use the MSTW2008 \cite{Martin:2009iq} sets of parton distributions (and QCD coupling) at each corresponding order.


In the first place we want to validate the use of the soft-virtual approximation, checking its accuracy at NLO, where the full result is known. 
We present the results for $d=3$ and $M_S=3.8\text{ TeV}$; we obtain similar results with the other sets of parameters.

\begin{figure}
\begin{center}
\begin{tabular}{c}
\epsfxsize=8.7truecm
\epsffile{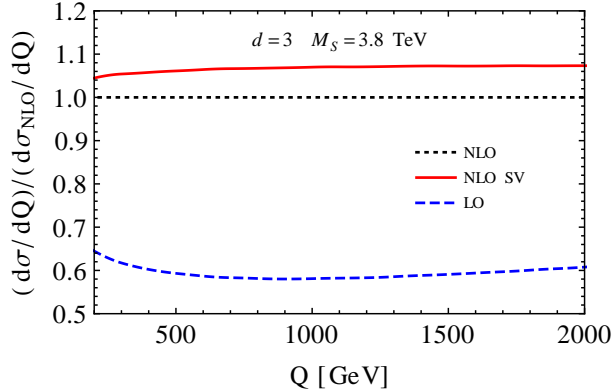}\\
\end{tabular}
\end{center}
\vspace{-0.7cm}
\caption{\label{NLOSVvsLO}
Ratio between the NLO-SV approximation and the full NLO result (red solid) compared with the ratio between the LO and the NLO cross sections (blue dashed) as a function of the di-lepton invariant mass.}
\end{figure}

In Figure \ref{NLOSVvsLO} we show the ratio between the approximation and the full NLO result as a function of the di-lepton invariant mass. We also show the ratio between the previous order (LO) and the NLO cross section. 
We can observe that the soft-virtual approximation reproduces very accurately the full result, with differences smaller than $10\%$.
Using the NLO-SV result clearly improves the accuracy of the prediction, since the previous order fails to reproduce the NLO by a $40\%$.
At NNLO we expect that the SV approximation will be even more accurate, since the size of the corrections is smaller.
Comparing with other processes dominated by gluon fusion in which both NNLO-SV and full NNLO have been computed, such as single \cite{Catani:2001ic,Harlander:2001is,Harlander:2002wh,Anastasiou:2002yz,Ravindran:2003um} and double \cite{deFlorian:2013uza,deFlorian:2013jea} Higgs production, we can expect differences with the exact NNLO result to be smaller than $5\%$.
We recall that the contribution of the gluon-gluon subprocess dominates the graviton production at the LHC in the di-lepton invariant mass region of the current analysis. 
For instance, at LO it contributes with $73\%$ of the cross section integrated between $Q=200\text{ GeV}$ and $Q=2000\text{ GeV}$.


With respect to the theoretical uncertainty, for the total cross section in the range $200\text{ GeV}\leq Q \leq 2000\text{ GeV}$ we find a scale variation close to $11\%$ at NLO, while in the case of the NLO-SV this value is about $5\%$, so that at this order the approximation underestimates the uncertainty by a factor $2$.

Once we have checked the validity of the approximation, we continue with the NNLO predictions.
We recall that our NNLO results are computed using the exact NLO cross section, and then adding the soft-virtual approximation only for the NNLO gravity corrections. For the SM contributions we use the exact NNLO result.
For simplicity, we will denote this computation as NNLO.

\begin{figure}
\begin{center}
\begin{tabular}{c}
\epsfxsize=10.0truecm
\epsffile{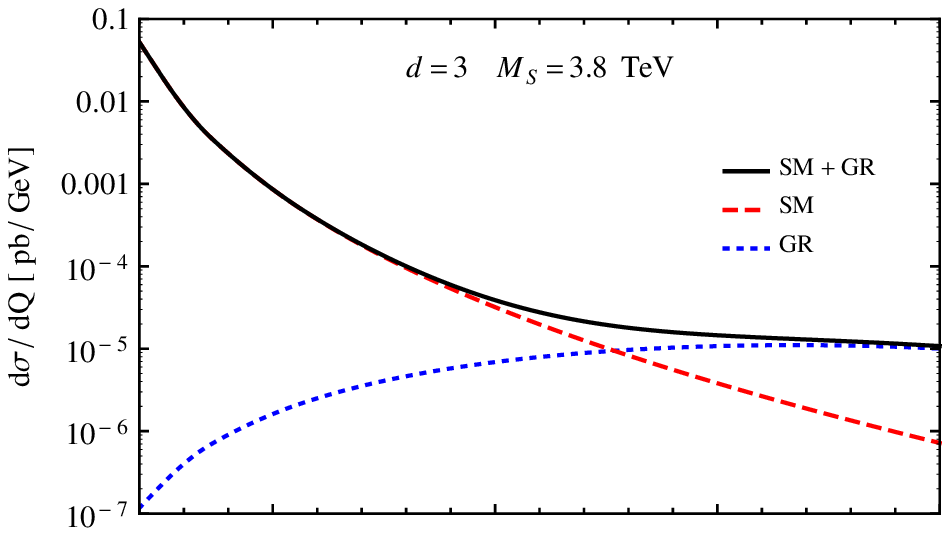}
\vspace{-0.7cm}
\\
\epsfxsize=9.45truecm
\hspace{0.07cm}
\epsffile{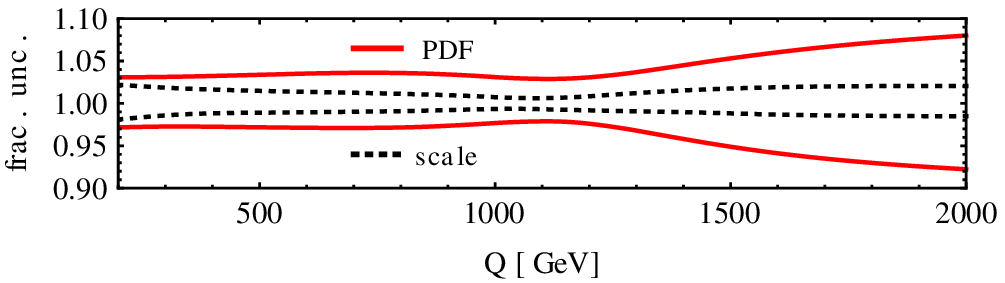}\\
\end{tabular}
\end{center}
\vspace{-0.6cm}
\caption{\label{SM_GR_Qdistr}
Di-lepton invariant mass distribution at the LHC ($\sqrt{s_H}=14\text{ TeV}$) for SM (blue-dotted), gravity (red-dashed) and SM+GR (black-solid) at NNLO. The lower inset gives the fractional scale (black-dotted) and PDF (red-solid) uncertainties.}
\end{figure}

In Figure \ref{SM_GR_Qdistr} we show the di-lepton invariant mass distribution for SM, GR and SM+GR at NNLO.
Deviations from the SM prediction can be observed for $Q\apgt 1000\text{ GeV}$.
For $Q\simeq 1200\text{ GeV}$, the SM and gravity contributions are of the same order, while for larger values of invariant mass the graviton mediated processes dominate the cross section.

We have also considered two different sources of theoretical uncertainties: missing higher orders in the QCD perturbative expansion and uncertainties in the determination of the parton flux.
To evaluate the size of the former we vary independently the factorization and renormalization scales in the range $0.5\,Q\leq\mu_F,\mu_R\leq2\,Q$, with the constraint $0.5\leq\mu_F/\mu_R\leq2$.
With respect to the PDFs uncertainties, we use the $90\%$ C.L. MSTW2008 sets \cite{Martin:2009iq}.
As we can observe from Figure \ref{SM_GR_Qdistr} the total scale variation is of ${\cal O}(5\%)$ in the whole range of invariant mass.
On the other hand, the PDF uncertainty is larger, specially in the gravity dominated invariant mass region, with a total variation close to $15\%$.
This different behaviour for small and large values of invariant mass originates from the larger fractional uncertainty of the gluon-gluon contribution (which dominates the graviton production) compared with the quark-antiquark one (which dominates the SM contribution).

\begin{figure}
\begin{center}
\begin{tabular}{c}
\epsfxsize=10.0truecm
\epsffile{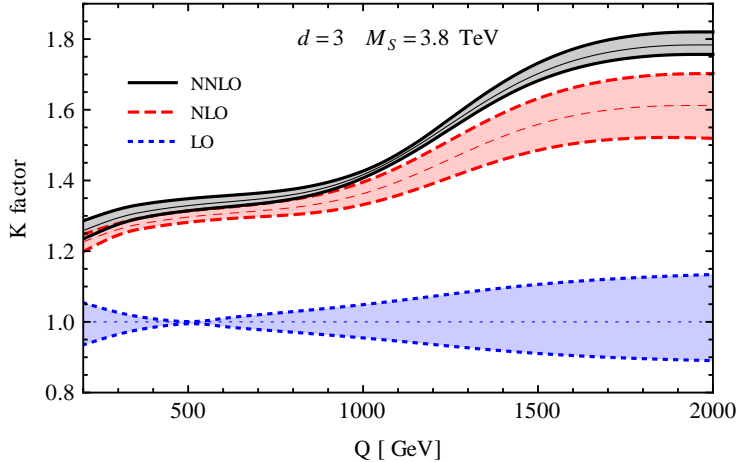} 
\\
\end{tabular}
\end{center}
\vspace{-0.7cm}
\caption{\label{kfactor}
$K$ factors as a function of the di-lepton invariant mass. The bands are obtained by varying the factorization and renormalization scales as indicated in the main text. The different curves correspond to the LO (blue-dotted), NLO (red-dashed) and NNLO (black-solid) predictions.}
\end{figure}

To evaluate the impact of the NNLO corrections we show in Figure \ref{kfactor} the corresponding $K$ factor as a function of the di-lepton invariant mass.
To normalize we use the LO prediction for $\mu_R=\mu_F=Q$.
The bands are obtained by varying the factorization and renormalization scales as indicated before.
We also include in the plot the previous order results.

We can observe, both at NLO and NNLO, the transition between the SM and the gravity dominated regions, $Q\aplt 1000\text{ GeV}$ and $Q\apgt 1000\text{ GeV}$ respectively.
Given that the QCD corrections for the graviton mediated di-lepton production are more sizeable than those of the SM Drell-Yan process, the NNLO $K$ factor goes from $K\simeq 1.3$ to $K\simeq 1.8$ as the value of $Q$ increases.
We can also see that there is an overlap between the NLO and NNLO bands for the small invariant mass region, while this does not happen for $Q\apgt 1000\text{ GeV}$.
This might be an effect due to the SV approximation if the underestimation of the uncertainty observed at NLO also holds at NNLO,
and we can expect the bands in the gravity dominated region to be larger in the exact NNLO result.
However, we also have to consider that an important part of the NNLO scale variation comes from the NLO contribution, for which we use the exact result.
At the same time, a small overestimation of the size of the NNLO corrections by the SV approximation (as it was observed at NLO in Figure \ref{NLOSVvsLO}) could be also contributing to this gap between the NLO and NNLO predictions.

\begin{figure}[t!]
\begin{center}
\begin{tabular}{c c}
\epsfxsize=7.0truecm
\epsffile{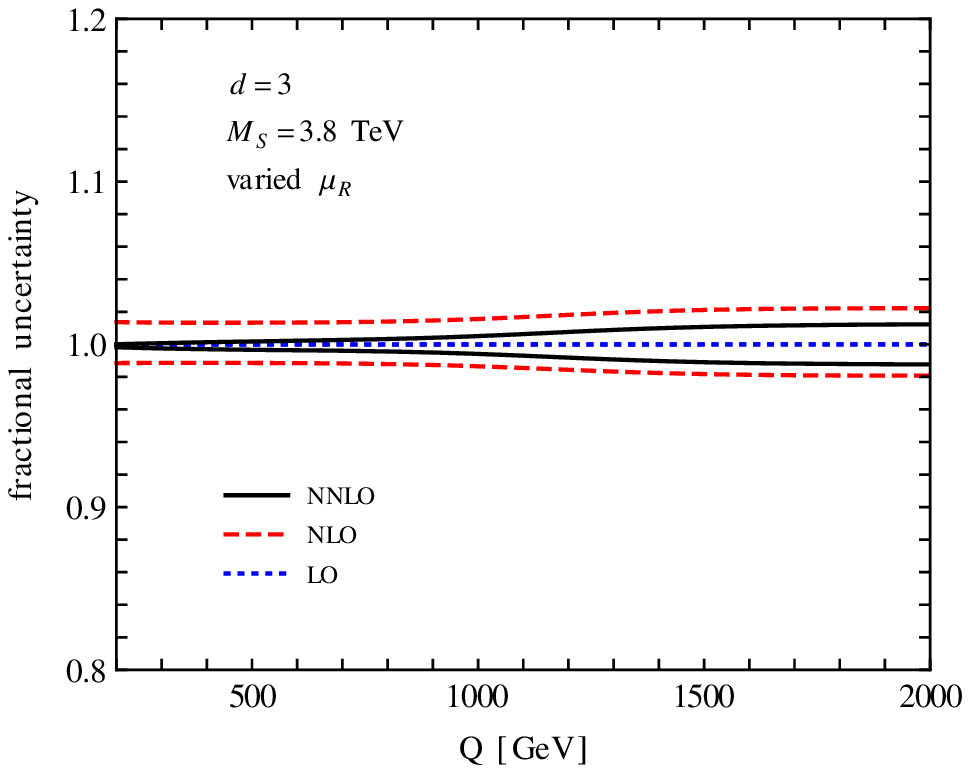} & 
\epsfxsize=7.0truecm
\epsffile{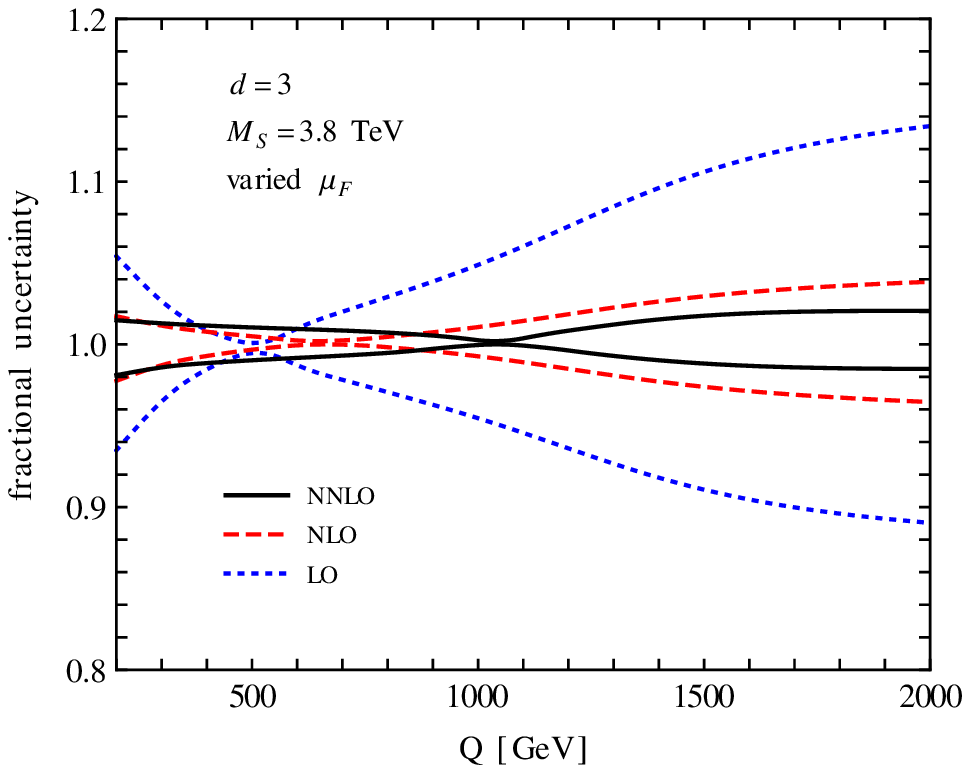} \\
\epsfxsize=7.0truecm
\epsffile{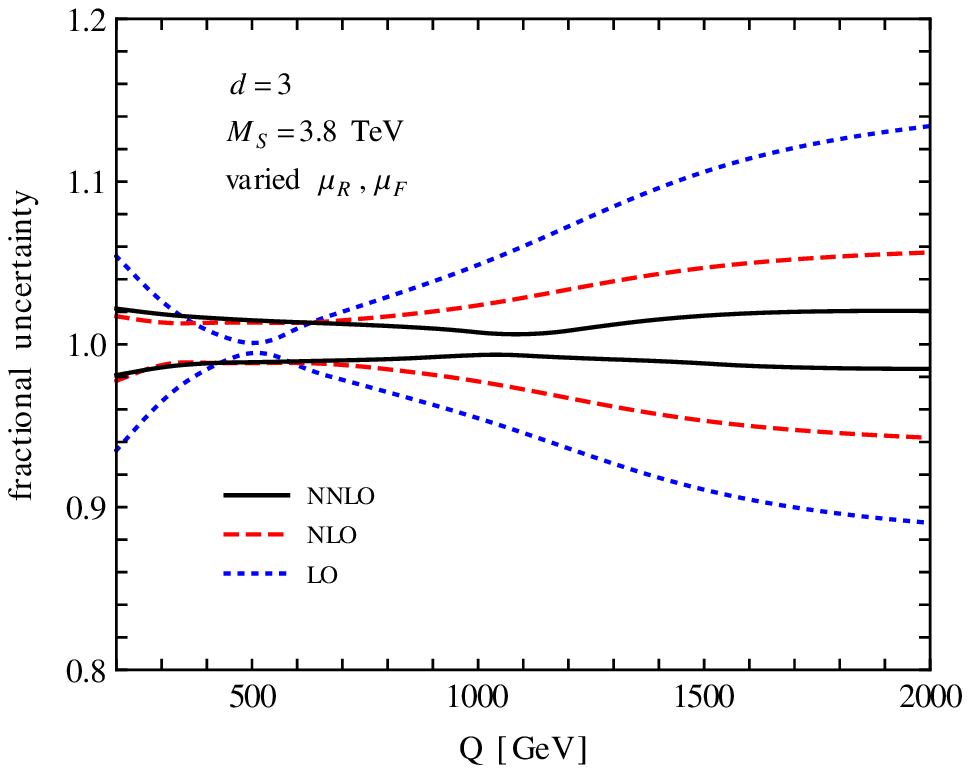} & 
\epsfxsize=7.0truecm
\epsffile{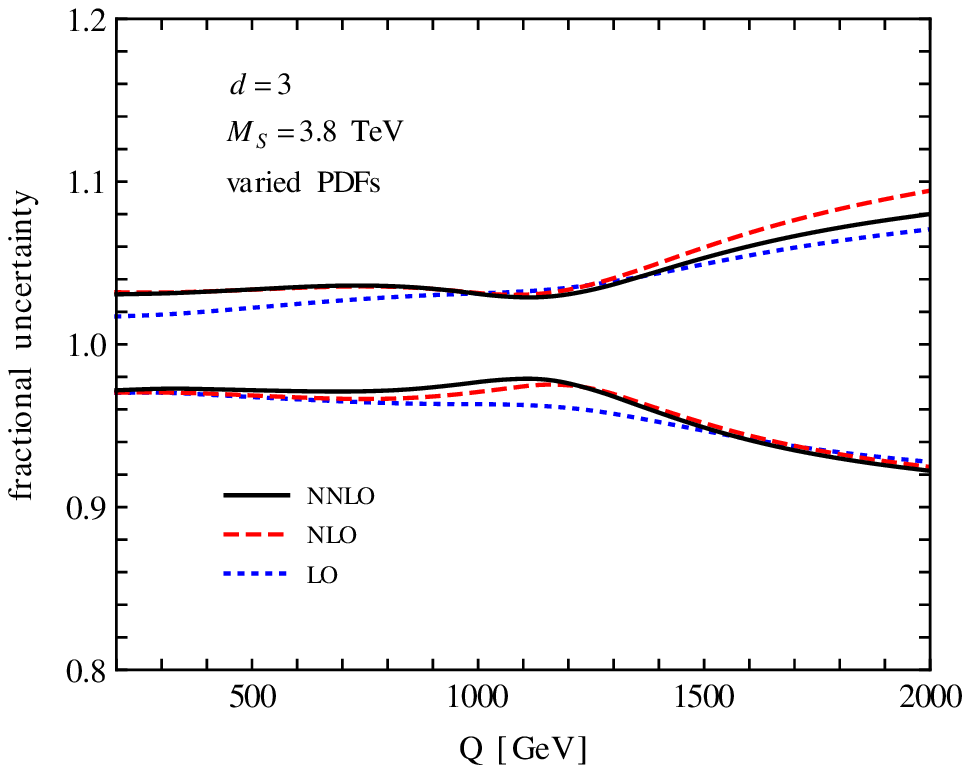} 
\\
\end{tabular}
\end{center}
\vspace{-0.7cm}
\caption{\label{uncertainties}
Fractional uncertainties of the di-lepton invariant mass distribution coming from $\mu_R$ variation (upper-left), $\mu_F$ variation (upper-right), $\mu_R, \mu_F$ variation (down-left) and PDF uncertainties (down right). In all cases we show the LO (blue-dotted), NLO (red-dashed) and NNLO (black-solid) predictions.}
\end{figure}

In Figure \ref{uncertainties} we present a more detailed analysis of the the theoretical uncertainties.
In the upper-left plot we show the fractional variation of the differential cross section as we vary the renormalization scale in the range $0.5Q\leq \mu_R\leq 2Q$, keeping $\mu_F = Q$.
Similarly, in the upper-right plot we vary $\mu_F$ keeping $\mu_R$ fixed.
Finally, in the down-left figure we show the total scale variation, varying simultaneously and independently both scales as indicated before.
On the other hand, in the down-right plot we present the fractional variation of the cross section coming from the parton flux determination uncertainties.
In all cases we show the LO, NLO and NNLO results.

We can observe that the $\mu_R$ dependence starts at NLO, with a total variation going from $3\%$ at $Q=200\text{ GeV}$ to $4\%$ at $Q=2000\text{ GeV}$.
At NNLO, the uncertainty is substantially reduced for the lower values of invariant mass, with a variation of less than $0.5\%$, while in the gravity dominated region the reduction is less significant.

As to the $\mu_F$ dependence, we can see that there is a zone of minimal variation which tends to move to higher values of invariant mass as we increase the order of the calculation.
Aside from that, in the large invariant mass region we can clearly observe how the uncertainty is reduced from LO to NLO and from NLO to NNLO.

The reduction of the uncertainties can be better observed in the total scale variation plot. As mentioned before, we can see that the NNLO total scale uncertainty remains quite constant in the whole range of invariant mass, with a value close to $4\%$.
As we can see from the plot, this result is three times smaller than the previous order uncertainty in the gravity dominated region.
On the other hand, for the SM dominated invariant mass region the NLO and NNLO scale variation is of the same order. 

Finally, we have the parton flux uncertainties, which as stressed before are the main source of theoretical uncertainties at NNLO.
In this case, we can observe that there is no significant difference between the results as we increase the order of the perturbative calculation.

\begin{figure}[t!]
\begin{center}
\begin{tabular}{c c}
\epsfxsize=8.0truecm
\epsffile{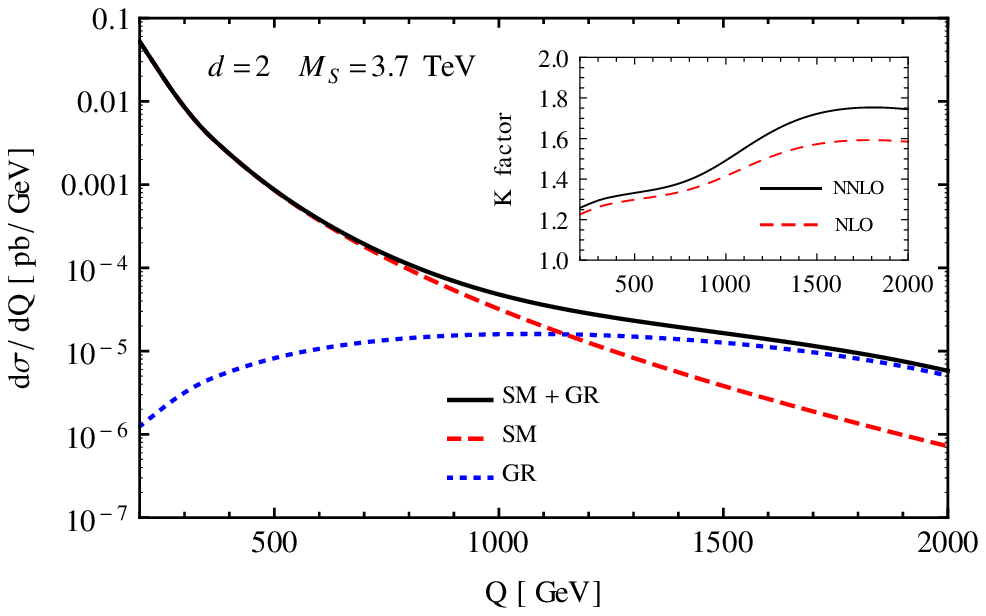} & 
\epsfxsize=8.0truecm
\epsffile{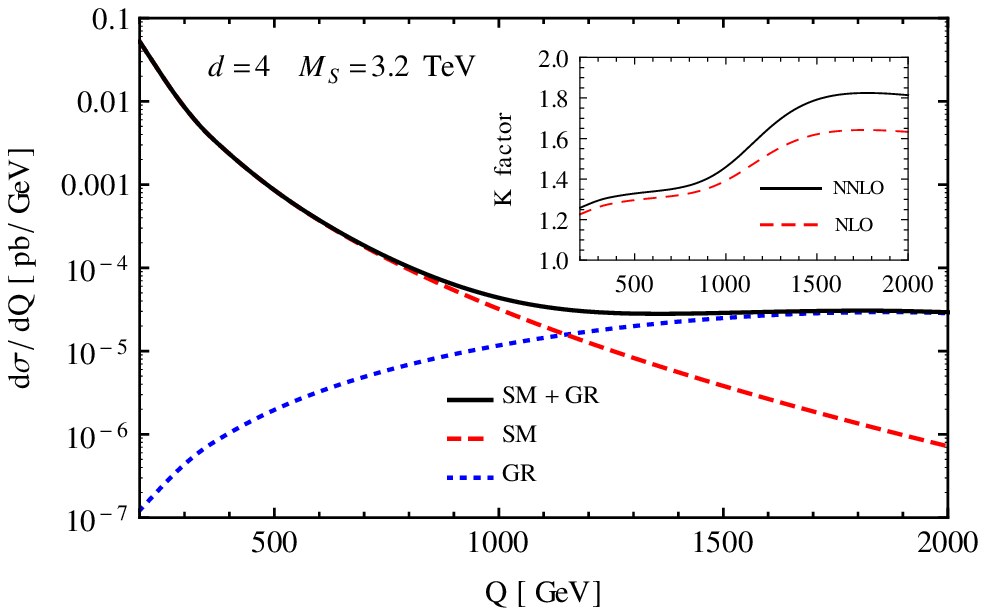} \\
\epsfxsize=8.0truecm
\epsffile{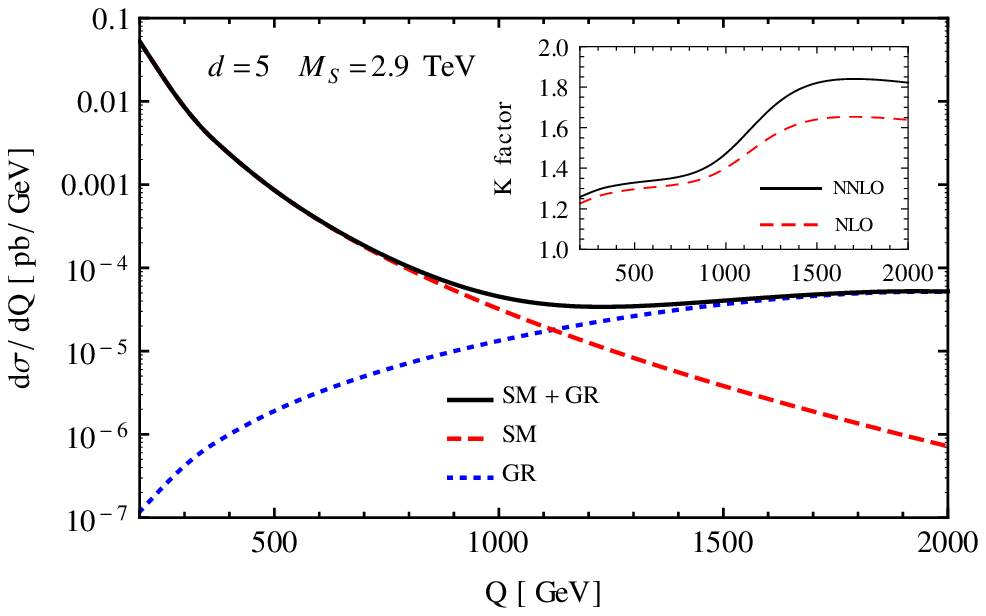} & 
\epsfxsize=8.0truecm
\epsffile{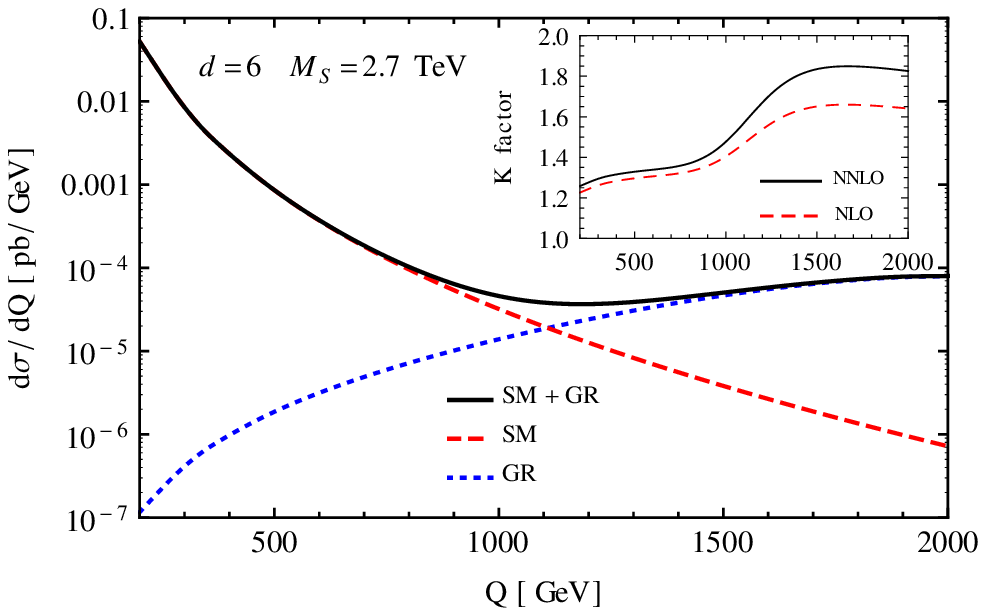} 
\\
\end{tabular}
\end{center}
\vspace{-0.7cm}
\caption{\label{parameters}
Di-lepton invariant mass distribution for SM (blue-dotted), gravity (red-dashed) and SM+GR (black-solid) at NNLO for $M_S=3.7\text{ TeV}$ and $d=2$ (upper-left), $M_S=3.2\text{ TeV}$ and $d=4$ (upper-right), $M_S=2.9\text{ TeV}$ and $d=5$ (down-left) and $M_S=2.7\text{ TeV}$ and $d=6$ (down-right). The inset plots show the corresponding $K$ factors at NLO (red-dashed) and NNLO (black-solid).}
\end{figure}

All the analysis described in this section was repeated for each of the model parameter sets, obtaining similar results.
In Figure \ref{parameters} we show the di-lepton invariant mass distributions for each of them at NNLO and the corresponding $K$ factors, for $\mu_F=\mu_R=Q$.
In all cases we can observe the same transition from the SM- to the GR-dominated region, and the resulting increase in the $K$ factor.

\subsection{RS Model}

We present now the predictions for the single graviton production in the Randall-Sundrum model at the LHC. Taking into account the latest bounds obtained by ATLAS \cite{ATLAS:2011ab} and CMS \cite{Chatrchyan:2011fq}, and the requirement $\Lambda_\pi\aplt 10\text{ TeV}$
, we have for each value of $\tilde{k}=k/{\overline M}_P$ a minimum and a maximum value of $M_1$ allowed.
At the same time, precision electroweak data and perturbativity requirements constrain the value of $\tilde{k}$ in the range $0.01\aplt \tilde{k} \aplt 0.1$
(some of these values are already excluded by the experiments).
In Table \ref{table_RS} we show the values of $\tilde{k}$ we used, and the corresponding minimum and maximum for $M_1$. These values explore the whole space of allowed parameters.

\begin{table}[b!]
\begin{center}
\begin{tabular}{l  c  c  c  c  c  c  c}
\hline\hline
$\tilde{k}$ & $0.04$ & $0.05$ & $0.06$ & $0.07$ & $0.08$ & $0.09$ & $0.1$ \\
\hline
$M_1^{\text{min}}[\text{TeV}]$ & $1.35$ & $1.55$ & $1.55$ & $1.65$ & $1.7$ & $1.8$ & $1.95$ \\
$M_1^{\text{max}}[\text{TeV}]$ & $1.55$ & $1.95$ & $2.3$  &  $2.7$ & $3.1$ & $3.45$ & $3.85$ \\
\hline\hline
\end{tabular}
\caption{Values of $\tilde{k}$ and $M_1$ used for the present analysis. \label{table_RS}}
\end{center}
\end{table}

\begin{figure}
\begin{center}
\begin{tabular}{c}
\epsfxsize=10.0truecm
\epsffile{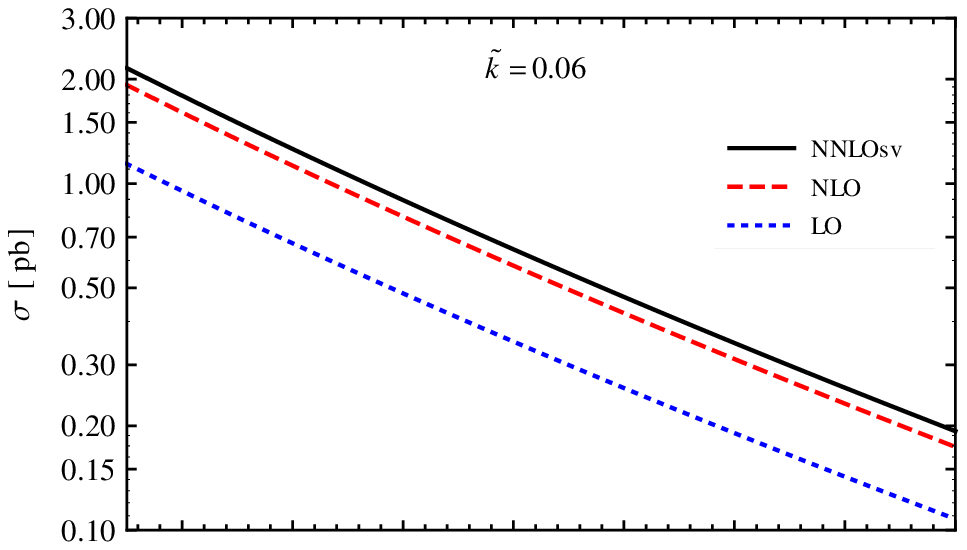}
\vspace{-0.7cm}
\\
\epsfxsize=9.65truecm
\hspace{-0.04cm}
\epsffile{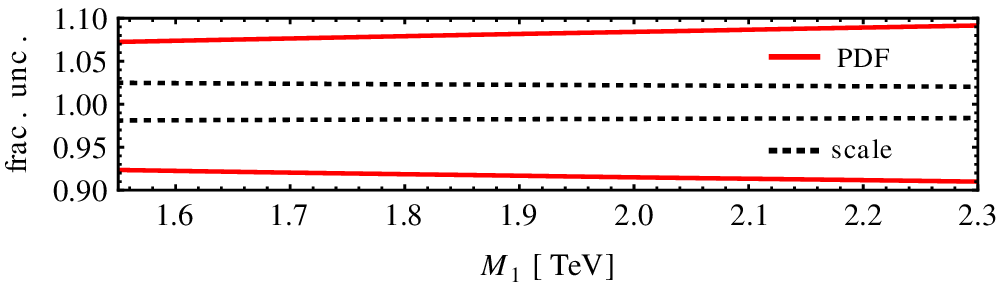}\\
\end{tabular}
\end{center}
\vspace{-0.6cm}
\caption{\label{RSxs}
Total single graviton production cross section at the LHC ($\sqrt{s_H}=14\text{ TeV}$) as a function of the lightest RS graviton mass at LO (blue-dotted), NLO (red-dashed) and NNLO (black-solid), the latest within the soft-virtual approximation. The lower inset gives the fractional scale (black-dotted) and PDF (red-solid) uncertainties.}
\end{figure}

In Figure \ref{RSxs} we show the total cross section as a function of the lightest RS graviton mass for $\tilde{k}=0.06$ at LO, NLO and NNLO, the latest within the soft-virtual approximation.
We can observe the exponential decay as we go to larger values of $M_1$.
The lower inset gives the fractional scale and PDF uncertainties.
We can observe that the scale variation remains almost constant throughout all the range of masses, with a total uncertainty of less than $5\%$.
Again, we could expect the exact NNLO uncertainty to be larger.
The PDF uncertainty is considerably larger, with a variation close to $15\%$ or $20\%$, depending on the value of $M_1$.

\begin{figure}
\begin{center}
\begin{tabular}{c}
\epsfxsize=10.0truecm
\epsffile{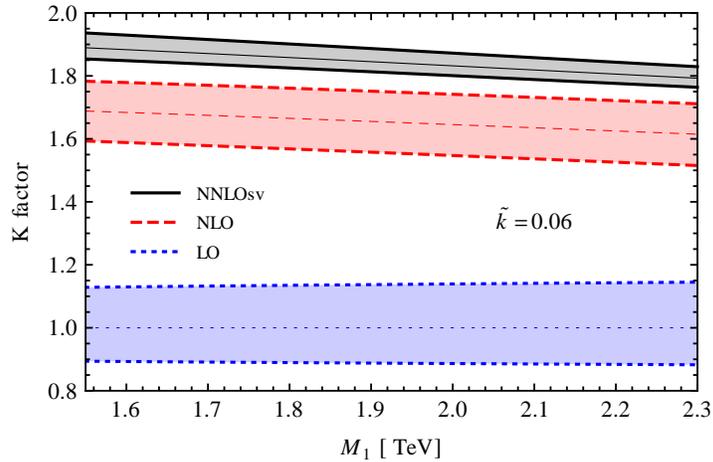} 
\\
\end{tabular}
\end{center}
\vspace{-0.7cm}
\caption{\label{KfacRS}
$K$ factors as a function of the lightest RS graviton. The bands are obtained by varying the factorization and renormalization scales as indicated in the main text. The different curves correspond to the LO (blue-dotted), NLO (red-dashed) and NNLO (black-solid) predictions, the last within the soft-virtual approximation.}
\end{figure}

The NNLO corrections are sizeable. This can be better seen in Figure \ref{KfacRS}, where we show the corresponding $K$ factor, again as a function of $M_1$.
We can observe that the $K$ factor is close to $1.9$ for the minimum value of $M_1$, and goes down to $1.8$ as we reach the maximum.
This represents an increase close to $15\%$ with respect to the NLO result.
We can also notice that the size of the bands, obtained performing the scale variation as indicated before, is considerably smaller at NNLO than in the previous orders.

Given that, for a fixed value of $M_1$, the size of $\tilde{k}$ only represents an overall normalization, the $K$ factor only depends on $M_1$.
We provide then the following analytic expression that parametrizes the NNLO-SV $K$ factor:
\beq
K_{NNLO}^{SV}=
2.207-0.239\left(\f{M_1}{1\text{ TeV}}\right)^{0.663}\,.
\eeq
This expression is valid for $1.35\text{ TeV}\leq M_1 \leq 3.85\text{ TeV}$, which includes the whole range of allowed values of $M_1$. The difference between this analytic expression and the exact NNLO-SV result is always smaller than $0.5\%$.
We remark that this expression is valid for any value of $\tilde{k}$ or $\Lambda_\pi$.


\section{Conclusions}

We have calculated the NNLO QCD corrections to the graviton production in models of TeV-scale gravity, working within the soft-virtual approximation, which is known to be very accurate for similar processes.
We expect that the differences between our predictions and the exact NNLO result will be smaller than $5\%$.

We considered the ADD and RS models. For the first, we computed the graviton contribution to the Drell-Yan process, while for the RS model we calculated the single graviton production cross section.

For the ADD model at the LHC, with a center-of-mass energy $\sqrt{s_H}=14\text{ TeV}$, we found a large $K$ factor ($K\simeq1.8$) for large values of the di-lepton invariant mass.
This region is dominated by the graviton contribution, whose QCD corrections are substantially larger than the SM ones.
The increment with respect to the previous order result is larger than $10\%$.

We also observe a substantial reduction in the scale uncertainty, with a total variation close to $4\%$. This value is about three times smaller than the NLO result in the large invariant mass region. Since at NLO the soft-virtual approximation underestimates the total uncertainty by a factor 2,
we can expect the exact NNLO scale variation to be larger. However, given that in our approximation the NLO contribution is treated in an exact way, and given that the NLO is an important contribution to the total NNLO variation, we can also expect the SV approximation to be more accurate at this order with respect to this source of theoretical uncertainty.
On the other hand, for the PDF uncertainty we found a total variation similar to what was found at NLO.

For the RS model we found a similar behaviour with respect to the NNLO QCD corrections.
In this case, we also provide a simple analytic parametrization of the NNLO $K$ factor, which only depends on $M_1$, and is valid for any value of $\tilde{k}$ or $\Lambda_\pi$.
Its value goes from $1.92$ for $M_1=1.35\text{ TeV}$
to $1.62$ for $M_1=3.85\text{ TeV}$.


\section*{Acknowledgements}

This work was supported in part by UBACYT, CONICET, ANPCyT and the Research Executive Agency (REA) of the European Union under the Grant Agreement number PITN-GA-2010-264564 (LHCPhenoNet).





\begin{thebibliography}{90}

\bibitem{ADD}
N.\ Arkani-Hamed, S.\ Dimopoulos and G.\ Dvali, Phys.\ Lett.\ B 429 (1998)
263; I.\ Antoniadis, N.\ Arkani-Hamed, S.\ Dimopoulos and G.\ Dvali,
Phys.\ Lett.\ B 436 (1998) 257; N.\ Arkani-Hamed, S.\ Dimopoulos and G.\
Dvali, Phys.\ Rev.\ D59 (1999) 086004.

\bibitem{RS1}
L.\ Randall and R.\ Sundrum, Phys. Rev. Lett. 83 (1999) 3370.

\bibitem{HLZ}
T.\ Han, J.\ D.\ Lykken and R.\ J.\ Zhang, Phys.\ Rev.\ D59 (1999) 105006

\bibitem{GRW}
G.\ F.\ Giudice, R.\ Rattazzi, and J.\ D.\ Wells, Nucl.\ Phys.\ B544 (1999) 3.

\bibitem{di-final}
J.\ L.\ Hewett, Phys.\ Rev.\ Lett.\ 82 (1999) 4765;
Prakash Mathews, Sreerup Raychaudhuri, K.\ Sridhar, 
Phys.\ Lett.\  B450 (1999) 343; JHEP 0007 (2000) 008.

\bibitem{tri-final}
M.C. Kumar, Prakash Mathews, V. Ravindran, Satyajit Seth, Phys.Rev. D85 (2012)
094507;
Li Xiao-Zhou, Duan Peng-Fei, Ma Wen-Gan, Zhang Ren-You, Guo Lei, 
Phys.\ Rev.\ D86 (2012) 095008.

\bibitem{di-ll1}
P.\ Mathews, V.\ Ravindran, K.\ Sridhar and W.\ L.\ van
Neerven, Nucl.\ Phys.\ B713 (2005) 333.
\bibitem{di-ll2}
P.\ Mathews, V. Ravindran, Nucl.\ Phys.\ B753 (2006) 1.
\bibitem{di-ll3}
M.C. Kumar, P.\ Mathews, V. Ravindran, Eur.\ Phys.\ J.\ C49 (2007) 599.

\bibitem{di-ph1}
M.C. Kumar, P.\ Mathews, V. Ravindran, A.\ Tripathi,
Phys.\ Lett.\ B672 (2009) 45.
%
\bibitem{di-ph2}
M.C. Kumar, Prakash Mathews, V.\ Ravindran, Anurag Tripathi,
Nucl.\ Phys.\ B818 (2009) 28.

\bibitem{di-ZZ1}
N. Agarwal, V. Ravindran, V. K. Tiwari, and A. Tripathi,
Nucl. Phys. B 830, 248 (2010).

\bibitem{di-ZZ2}
 N.~Agarwal, V.~Ravindran, V.~K.~Tiwari and A.~Tripathi,
  Phys.\ Lett.\ B {\bf 686} (2010) 244
  [arXiv:0910.1551 [hep-ph]].

\bibitem{di-WW1}
N.\ Agarwal, V.\ Ravindran, V.\ K.\ Tiwari and A.\ Tripathi, Phys.\ Rev.\ D82
(2010) 036001.

\bibitem{di-WW2}
N.\ Agarwal, V.\ Ravindran, V.\ K.\ Tiwari and A.\ Tripathi, Phys.\ Lett.\ B 690
(2010) 390.

\bibitem{di-ph+ps}
R.\ Frederix, 
M.\ K.\ Mandal, P.\ Mathews, V.\ Ravindran, S.\ Seth, P.\ Torrielli and M.\ Zaro, 
JHEP 1212 (2012) 102.

\bibitem{di-final+ps}
R.\ Frederix, M.\ K.\ Mandal, P.\ Mathews, V.\ Ravindran, S.\ Seth, 
arXiv:1307.7013.

\bibitem{jEt}
S. Karg, M. Karamer, Q. Li, and D. Zeppenfeld, Phys. Rev. D 81, 094036 (2010).

\bibitem{phEt}
X. Gao, C. S. Li, J. Gao, and J. Wang, Phys. Rev. D 81, 036008 (2010).

\bibitem{ZWEt}
M. C. Kumar, P. Mathews, V. Ravindran, and S. Seth, Nucl. Phys. B847, 54 (2011);
J. Phys. G 38, 055001 (2011).



\bibitem{deFlorian:2013sza}
  D.~de Florian, M.~Mahakhud, P.~Mathews, J.~Mazzitelli and V.~Ravindran,
  arXiv:1312.6528 [hep-ph].



\bibitem{deFlorian:2012za}
  D.~de Florian and J.~Mazzitelli,
  JHEP {\bf 1212} (2012) 088
  [arXiv:1209.0673 [hep-ph]].
  
  \bibitem{Ravindran:2005vv}
  V.~Ravindran,
  Nucl.\ Phys.\ B {\bf 746} (2006) 58
  [hep-ph/0512249].
  
\bibitem{Ravindran:2006cg}
  V.~Ravindran,
  Nucl.\ Phys.\ B {\bf 752} (2006) 173
  [hep-ph/0603041].

\bibitem{Moch:2005ky}
  S.~Moch and A.~Vogt,
  Phys.\ Lett.\ B {\bf 631} (2005) 48
  [hep-ph/0508265].
  
\bibitem{Laenen:2005uz}
  E.~Laenen and L.~Magnea,
  Phys.\ Lett.\ B {\bf 632} (2006) 270
  [hep-ph/0508284].
  
\bibitem{Idilbi:2005ni}
  A.~Idilbi, X.~-d.~Ji, J.~-P.~Ma and F.~Yuan,
  Phys.\ Rev.\ D {\bf 73} (2006) 077501
  [hep-ph/0509294].
  
\bibitem{Catani:2003zt}
  S.~Catani, D.~de Florian, M.~Grazzini and P.~Nason,
  JHEP {\bf 0307} (2003) 028
  [hep-ph/0306211].

\bibitem{Hamberg:1990np}
  R.~Hamberg, W.~L.~van Neerven and T.~Matsuura,
  Nucl.\ Phys.\ B {\bf 359} (1991) 343
   [Erratum-ibid.\ B {\bf 644} (2002) 403].
  
\bibitem{Harlander:2002wh}
  R.~V.~Harlander and W.~B.~Kilgore,
  Phys.\ Rev.\ Lett.\  {\bf 88} (2002) 201801
  [hep-ph/0201206].



  
\bibitem{Catani:1996yz}
  S.~Catani, M.~L.~Mangano, P.~Nason and L.~Trentadue,
  Nucl.\ Phys.\ B {\bf 478} (1996) 273
  [hep-ph/9604351].
  

  
\bibitem{ATLAS:2011ab}
  G.~Aad {\it et al.}  [ATLAS Collaboration],
  Phys.\ Lett.\ B {\bf 710} (2012) 538
  [arXiv:1112.2194 [hep-ex]].

\bibitem{Chatrchyan:2011fq}
  S.~Chatrchyan {\it et al.}  [CMS Collaboration],
  Phys.\ Rev.\ Lett.\  {\bf 108} (2012) 111801
  [arXiv:1112.0688 [hep-ex]].


\bibitem{Martin:2009iq}
  A.~D.~Martin, W.~J.~Stirling, R.~S.~Thorne and G.~Watt,
  Eur.\ Phys.\ J.\ C {\bf 63} (2009) 189
  [arXiv:0901.0002 [hep-ph]].
  
\bibitem{Catani:2001ic} 
  S.~Catani, D.~de Florian and M.~Grazzini,
  JHEP {\bf 0105}, 025 (2001)
  [hep-ph/0102227].
  
\bibitem{Harlander:2001is}
  R.~V.~Harlander and W.~B.~Kilgore,
  Phys.\ Rev.\ D {\bf 64} (2001) 013015
  [hep-ph/0102241].
  
\bibitem{Harlander:2002wh}
  R.~V.~Harlander and W.~B.~Kilgore,
  Phys.\ Rev.\ Lett.\  {\bf 88} (2002) 201801.
[arXiv:hep-ph/0201206].

\bibitem{Anastasiou:2002yz}
  C.~Anastasiou and K.~Melnikov,
  Nucl.\ Phys.\  B {\bf 646} (2002) 220.
[arXiv:hep-ph/0207004].

\bibitem{Ravindran:2003um}
  V.~Ravindran, J.~Smith and W.~L.~van Neerven,
  Nucl.\ Phys.\  B {\bf 665} (2003) 325.
[arXiv:hep-ph/0302135].

\bibitem{deFlorian:2013uza}
  D.~de Florian and J.~Mazzitelli,
  Phys.\ Lett.\ B {\bf 724} (2013) 306
  [arXiv:1305.5206 [hep-ph]].
  
\bibitem{deFlorian:2013jea}
  D.~de Florian and J.~Mazzitelli,
   Phys.\ Rev.\ Lett.\  {\bf 111} (2013) 201801
  [arXiv:1309.6594 [hep-ph]].


\end{thebibliography}
\end{document}